\documentclass[10pt,aps,prd,twocolumn,floats,floatfix,showpacs,superscriptaddress,nofootinbib]{revtex4-1}

\usepackage{fancyhdr}
\usepackage{amsmath,amsthm,amssymb}
\usepackage{enumerate}
\usepackage{color}
\usepackage{graphicx}
\usepackage{caption}

\def\e{\epsilon}
\def\E{\mathcal{E}}
\def\L{\mathcal{L}}
\def\G{\mathcal{G}}
\def\H{\mathcal{H}}

\def\beq{\begin{equation}}
\def\eeq{\end{equation}}
\def\bea{\begin{eqnarray}}
\def\eea{\end{eqnarray}}

\def\bwt{\begin{widetext}}
\def\ewt{\end{widetext}}

\begin{document}
\author{Mehdi Saravani}
\affiliation{School of Mathematical Sciences, University of Nottingham, University Park, Nottingham, NG7 2RD, UK}
\author{Thomas P. Sotiriou}
\affiliation{School of Mathematical Sciences, University of Nottingham, University Park, Nottingham, NG7 2RD, UK}
\affiliation{School of Physics and Astronomy, University of Nottingham, University Park, Nottingham, NG7 2RD, UK}

\title{Classification of shift-symmetric Horndeski theories and hairy black holes}

\begin{abstract}

No-hair theorems for scalar-tensor theories imply that the trivial scalar field configuration is the unique configuration around stationary black hole spacetimes. The most basic assumption in these theorems is that a constant scalar  configuration is actually admissible. 
In this paper, we classify shift-symmetric Horndeski theories according to whether or not they admit the trivial  scalar configuration as a solution and under which conditions. Local Lorentz symmetry and the presence of a  linear coupling between the scalar field and Gauss-Bonnet invariant plays feature prominently in this classification. We then use the classification to show that any theory without linear Gauss-Bonnet coupling that respects Local Lorentz symmetry admits all GR solutions. 
We also study the scalar hair configuration around black hole spacetimes in theories  where the linear Gauss-Bonnet coupling is present. We show that the scalar hair of the configuration is secondary, fixed by the regularity of the horizon, and is determined by the black hole horizon properties.  
\end{abstract}

\maketitle
\section{Introduction}
After more than 100 years of its introduction, General Relativity (GR) is still the most successful theory describing gravitational interactions. Despite its mathematical simplicity and consistency with observations, many alternative theories have been proposed with motivations ranging from quantum gravity, cosmology and resolving dark matter/energy to testing Lorentz violation in the gravity sector. 
Any modification to GR (in 3+1 dimensions), by Lovelock's theorem \cite{Lovelock:1971yv, Lovelock:1972vz}, introduces new degrees of freedom. In this regard, scalar-tensor theories, which possess an additional scalar field,  are perhaps the simplest alternatives to GR. 

The new scalar degree of freedom could lead to intriguing phenomenology in black hole spacetimes. No hair theorems \cite{1970CMaPh..19..276C, Hawking:1972qk,Bekenstein:1995un,Sotiriou:2011dz,Hui:2012qt,Sotiriou:2013qea,Sotiriou:2015pka,Silva:2017uqg} seem to suggest the opposite. They imply that, under certain assumptions regarding symmetries and asymptotics,  the trivial configuration (vanishing scalar field) is the unique acceptable configuration in a black hole spacetime. However, no-hair theorems can be circumvented by relaxing these assumption --- most notably allowing the scalar to be nonstationary --- or by exploiting instabilities. This has lead to various scenarios for long-lived scalar hair \cite{Arvanitaki:2010sy, Cardoso:2013opa, Cardoso:2013fwa, Babichev:2013cya, Herdeiro:2014goa, Babichev:2016rlq, Doneva:2017bvd, Antoniou:2017acq, Silva:2017uqg}.  

Perhaps the most obvious way to find theories for which black holes have hair is to violate the most basic assumption of no-hair theorems: that a trivial configuration for the scalar is admissible in the first place for black holes spacetimes. Indeed, in Ref.~\cite{Hui:2012qt} is was shown that static, spherically symmetric and asymptotically flat black holes cannot have hair in shift-symmetric scalar tensor theories. Shirt symmetry implies invariance under the transformation $\phi\to \phi+$constant and it can be thought of as the symmetry that prevents the scalar to acquire a mass from quantum corrections. 
Subsequently, it was pointed out in Ref.~\cite{Sotiriou:2013qea} that a linear coupling between a scalar and the Gauss-Bonnet invariant respect shift symmetry and at the same time leads to a contribution to the scalar's field equation that depends only on the Gauss--Bonnet invariant.\footnote{Beyond the confines of shift symmetry, it was well known that a coupling between the scalar and the Gauss-Bonnet invariant leads to black hole hair \cite{Kanti:1995vq,Yunes:2011we}. } The latter does not vanish in black hole spacetimes, it sources the scalar, and makes a constant scalar configuration inadmissible. It was further shown there that this is the only coupling term with this property, which at the same time allows for a constant scalar configuration in flat space (the the Gauss--Bonnet invariant vanished).  The latter is a requirement if one wants the theory to respect local Lorentz symmetry (LLS). The gradient of the scalar field $\nabla_a \phi$ picks a preferred direction at any point in spacetime. Hence, having a nontrivial gradient in flat space is a violation of LLS.

Let us now consider the implications of the results in Ref.~\cite{Sotiriou:2013qea} in terms of the shift-symmetric 
(SS) Horndeski theory \cite{1974IJTP...10..363H,Deffayet:2011gz, Kobayashi:2011nu}. This is the most general shift-symmetric scalar tensor theory which leads to second order equation of motion upon variation. The Lagrangian of the theory reads 
\begin{equation}\label{L_Horndeski}
	\L
	=
	\L_{2} + \L_{3} + \L_{4} + \L_{5}
\end{equation}
where
\begin{align}
	\L_{2} &= G_{2}, \\
	\L_{3} &= - G_{3} \Box \phi, \\
	\L_{4} &= G_{4} R + G_{4X} \left[ \left( \Box \phi \right)^{2} - \left( \nabla_{a} \nabla_{b} \phi \right)^{2} \right], \\
	\L_{5} &= G_{5} G_{ab} \nabla^{a} \nabla^{b} \phi \\ &- \frac{1}{6} G_{5X} \left[ \left( \Box \phi \right)^{3} - 3 \Box \phi \left( \nabla_{a} \nabla_{b} \phi \right)^{2} + 2 \left( \nabla_a \nabla_b \phi \right)^{3} \right],
\end{align}
$X = - \frac{1}{2} \nabla_{a} \phi \nabla^{a} \phi$ and $G_{iX}=\partial_{X} G_{i}$. $G_{i}$'s are functions of $\phi$ and $X$. Shift symmetry implies that  $G_i$'s only depend on $X$ \cite{Sotiriou:2014pfa}. (Throughout this paper we are assuming mostly positive signature for the metric.) According to Ref.~\cite{Sotiriou:2013qea}, the no-hair theorem of Ref.~\cite{Hui:2012qt} should be applicable to SS Horndeski theories provide that: (i) the term $\alpha \phi {\cal G}$, where $\alpha$ is a coupling constant and ${\cal G}\equiv R^{abcd}R_{abcd}-4 R^{ab}R_{ab}+R^2$ is the Gauss--Bonnet invariant, is entirely absent; (ii) the functions $G_i(X)$ are such that LLS is respected. It is worth pointing out that it is easy to find examples that do not respect these conditions, {e.g.}
\begin{enumerate}
\item[a)] Cuscuton theory \cite{Afshordi:2006ad} is a SS Horndeski theory given by $G_2=\sqrt{|X|}$. Deriving the equation of motion, one could directly check that the theory does not have $\phi = 0$ as a solution.
\item[b)] A massless scalar field with linear Gauss-Bonnet coupling
\beq\label{massless_GB}
\L=-\frac{1}{2}\nabla_a \phi \nabla^a \phi +\alpha \phi \mathcal{G}.
\eeq
This is a shift-symmetric theory as the Gauss-Bonnet term is a total divergence in four dimension, and thus belongs to SS Horndeski class. In terms of the $G_i$ functions, the theory is given by $G_2=X$ and $G_5=-4\alpha \ln|X|$ \cite{Kobayashi:2011nu}. The equation of motion from Lagrangian \eqref{massless_GB} reads
\beq
\Box \phi +\alpha \mathcal{G}=0,
\eeq
which clearly shows $\phi= 0$ is not in the space of solutions on a generic background spacetime \cite{Sotiriou:2013qea}. 
\end{enumerate}

%

The discussion above suggest strongly that one should be able to classify SS Horndeski theories according to whether they accept the trivial configuration $\phi =0$ as a solution in flat space and in general spacetimes. It also suggests that this classification could help understand the properties of black holes in these theories. Below we introduce this classification, we determine the condition that a theory needs to satisfy (in terms of the $G_i$ functions) in order to belong in a certain class, and we use it to uncover some interesting properties for the theories that belong in each class. We expect these properties, and the classification in general, to be useful in various applications. Sticking to our initial motivation, we use it to prove that black hole in Horndeski theory cannot have an independent charge.

\section{Classification}

In this section, we identify three mutually exclusive classes in SS Horndeski theories.
Let us define $\E_\phi[\phi,g]$ to be the scalar field equation of motion derived from Lagrangian \eqref{L_Horndeski}. Then we define:\\
{\bf Class 1}: SS Horndeski theories satisfying 
\beq\label{class1_condition}
\E_\phi[\phi=0,\,g]=0,\qquad \forall g.
\eeq
These theories admit $\phi=0$ solutions for the scalar and hence all of the solutions of GR.\\
{\bf Class 2}: SS Horndeski theories which are not in Class 1 but satisfying
\beq\label{class2_condition}
\lim_{g\rightarrow \eta}\E_\phi [\phi=0,\, g]=0,
\eeq
where $\eta_{ab}$ is the Minkowski metric. These theories do not admit $\phi=0$ solutions and hence they do not share GR solutions in general. However, they have Minkowski spacetime as a solution, and this solution is smoothly connected to other solutions.\\
{\bf Class 3}: This class is defined as the complement of Classes 1 and 2 combined, {\em i.e.} it consists of theories that either do not admit flat space with as a solution at all, or they formally admit it but it is a disconnected solution. 

Following the discussion in the introduction regarding Lorentz symmetry, theories in Classes 1 and 2 respect LLS , while Class 3 theories are Lorentz-violating.
Before ending this section, let us mention an important result regarding the scalar field equation of motion which we use extensively in the following sections. By virtue of shift symmetry, the theory acquires a Noether's current $J^a$ associated to the shift symmetry which is given explicitly by
\bea
J^a&=&G_{2X}J^a_{(2,1)}+G_{3X}J^a_{(3,1)}\notag\\
&&+G_{4X}J^a_{(4,1)}+G_{4XX}J^a_{(4,2)}\notag\\
&&+G_{5X}J^a_{(5,1)}+G_{5XX}J^a_{(5,2)},
\eea
where 
\bea
J^a_{(2,1)}&=&-\nabla^a \phi,\\
J^a_{(3,1)}&=&\nabla^a \phi \Box \phi-\nabla^a\nabla^b\phi\nabla_b\phi,\\
J^a_{(4,1)}&=&2G^{ab}\nabla_b\phi,\\
J^a_{(4,2)}&=&(\nabla_c\nabla_d\phi)^2\nabla^a\phi-(\Box\phi)^2\nabla^a\phi\notag\\
&+&2\Box\phi\nabla^a\nabla^b\phi\nabla_b\phi-2\nabla^b\nabla^c\phi\nabla_c\phi\nabla^a\nabla_b\phi,\\
J^a_{(5,1)}&=&-\nabla^a\phi G^{cd}\nabla_c\nabla_d\phi+G^{ab}\nabla_b\nabla_c\phi\nabla^c\phi-\Box\phi R^{ab}\nabla_b\phi\notag\\
&+&R^{cbda}\nabla_c\nabla_d\phi\nabla_b\phi+R^{cd}\nabla_c\phi\nabla^a\nabla_d\phi,\\
J^a_{(5,2)}&=&\frac{\nabla^a\phi}{6}\left[(\Box\phi)^3-3\Box\phi(\nabla_c\nabla_d\phi)^2+2(\nabla_c\nabla_d\phi)^3\right]\notag\\
&-&\frac{1}{2} (\Box\phi)^2\nabla^a\nabla^b\phi\nabla_b\phi+\frac{1}{2}(\nabla_c\nabla_d\phi)^2\nabla^a\nabla^b\phi\nabla_b\phi\notag\\
&+&\Box\phi\nabla^b\nabla^c\phi\nabla_c\phi\nabla^a\nabla_b\phi\notag\\
&-&\nabla^b\nabla^c\phi\nabla_c\phi\nabla^a\nabla^d\phi\nabla_d\nabla_b\phi.
\eea
The equation of motion for the scalar field can be thought of as the conservation of the Noether's current, namely
\beq
\E_\phi[\phi,\, g]=- \nabla_a J^a.
\eeq

Inspecting carefully the terms in the current shows that they have a specific scaling dimension with $\phi$, $J^a_i[\beta\phi]=\beta^{n_i}J^a_i[\phi]$, where
\bea
n_{(2,1)}=1,&&\qquad n_{(3,1)}=2,\notag\\
n_{(4,1)}=1,&&\qquad n_{(4,2)}=3,\notag\\
n_{(5,1)}=2,&&\qquad n_{(5,2)}=4.\label{scaling_n}
\eea
We will use this scaling property in the next sections. 

\subsection{Class 1: Theories with GR solutions}
In this section, we derive the conditions on the $G_i$'s for a theory to be in Class 1. We impose the defining condition of the class as follows
\beq\label{class1_condition_modified}
0=\E_\phi[\phi=0,\, g]=\lim_{\e\rightarrow 0}~ \E_\phi[\e \bar \phi,\,g] \qquad \forall g,
\eeq
where $\bar \phi$ can be any differentiable field configuration. This is to ensure that the $\phi=0$ solution can be reached smoothly. In other words, we disregard theories in which the GR branch is dynamically disconnected from any other solution. In order to make use of the scaling properties in eq. \eqref{scaling_n}, we define the following functions
\bea
F_{(2,1)}(X)=|X|^{1/2}G_{2X},&&\qquad F_{(3,1)}(X)=XG_{3X},\notag\\
F_{(4,1)}(X)=|X|^{1/2}G_{4X},&&\qquad F_{(4,2)}(X)=|X|^{3/2}G_{4XX},\notag\\
F_{(5,1)}(X)=XG_{5X},&&\qquad F_{(5,2)}(X)=X^2G_{5XX}
\eea
and currents 
\bea
j^a_{(2,1)}=|X|^{-1/2}J^a_2,&\qquad &j^a_{(3,1)}=X^{-1}J^a_3\notag\\
j^a_{(4,1)}=|X|^{-1/2}J^a_{(4,1)},&\qquad &j^a_{(4,2)}=|X|^{-3/2}J^a_{(4,2)}\notag\\
j^a_{(5,1)}=X^{-1}J^a_{(5,1)},&\qquad &j^a_{(5,2)}=X^{-2}J^a_{(5,2)}
\eea
and rewrite the Noether's current as
\bea
&J^a&=F_{(2,1)}j^a_{(2,1)}+F_{(3,1)}j^a_{(3,1)}\notag\\
&&+F_{(4,1)}j^a_{(4,1)}+F_{(4,2)}j^a_{(4,2)}\notag\\
&&+F_{(5,1)}j^a_{(5,1)}+F_{(5,2)}j^a_{(5,2)}.\label{current_scaling_def}
\eea
The scalings are designed such that the new currents $j$'s are scale invariant $j^a_{i}[\e \bar \phi,\,g]=j^a_{i} [\bar \phi,\,g]$; the $\e$-dependence is now encoded in $F_i$'s. In particular, the behaviour of $F_i(X)$ close to $X=0$ is crucial for $\e\rightarrow 0$ limit.

If $F_i$ functions are such that $F_i(X=0)=0$, then the current $J^a$ vanishes as $\epsilon \rightarrow0$.  Thus the equation of motion is satisfied. As a result, $F_i(X=0)=0$ are {\it sufficient} conditions for a theory to belong to Class 1. In what follows, we show that $F_i(X=0)=0$ are necessary as well.

According to our definition, a theory is in Class 1 if for all spacetime metrics $g_{ab}$
\beq
\lim_{\e \rightarrow 0}~\E_\phi[\e \bar \phi, \,g]=0.
\eeq
Let us restrict ourselves to static spherically symmetric field configuration $\bar \phi$ and metric and choose the metric to be infinitesimally close to the flat metric,
\beq
g_{ab}dx^a dx^b=-(1+\e_1 h(r))dt^2+\frac{dr^2}{1+\e_2f(r)}+r^2 d\Omega^2,
\eeq
where $\e_1$ and $\e_2$ are small numbers. With staticity, spherical symmetry and smoothness at the centre, the scalar field equation of motion reduces to $J^r=0$, thus we require 
\beq
\lim_{\e\rightarrow 0}~J^r[\e \bar \phi,\, g]=0.
\eeq
Imposing above to hold in any order of $\e_1$ and $\e_2$, we get at $X=0$
\bea
\e_1^0\e_2^0~\mbox{order: }&&F_{(2,1)}=F_{(3,1)}=F_{(4,2)}=0\\
\e_1~\mbox{and} ~\e_2~\mbox{order: }&&F_{(4,1)}=0,~F_{(5,1)}=-F_{(5,2)}\\
\e_1\e_2~\mbox{order: }&& 5F_{(5,1)}+4F_{(5,2)}=0.
\eea
The combination of above conditions yields 
\beq\label{class1_result}
F_i(X=0)=0.
\eeq
Eq. \eqref{class1_result} is derived by restricting the metric to be static, spherically symmetric and close to flat spacetime, thus it constitutes a set of {\it necessary} conditions for eq. \eqref{class1_condition_modified} to hold. Combining this with the previous result, we have shown that $F_i(X=0)=0$ is a set of necessary and sufficient conditions for a theory to be in Class 1. 

On a final note we should mention that all $F_i(X=0)$, if not divergent, are not independent since they are related through their definitions as derivatives of $G_i$ functions. In particular, 
\bea
F_{(4,2)}(X=0) &=& -\frac{1}{2} F_{(4,1)}(X=0),\notag\\
F_{(5,1)}(X=0) &=& - F_{(5,2)}(X=0).\label{F_relation}
\eea

\subsection{Class 2: Theories with Minkowski solution}
So far, we have focused on theories which admit $\phi=0$ as a solution on any background spacetime metric. Now, we turn our attention to Class 2 theories which do not satisfy this property and they only admit the trivial solution on a flat spacetime. 

In the definition of Class 2 in eq.~\eqref{class2_condition}, we require the limit $g_{ab}\rightarrow \eta_{ab}$ to exclude theories that have $\phi=0$ solution on flat spacetime when this solution cannot be reached smoothly. Intuitively, eq.~\eqref{class2_condition} requires that, starting from a perturbed flat metric and then damping the perturbations, there is always a solution to the scalar field equation of motion which is continuously connected to $\phi=0$.

Considering a static spherically symmetric metric and taking the limit to the flat metric in eq. \eqref{class2_condition}, similar to our argument in the previous section, we get the following {\it necessary} conditions at $X=0$
\bea
&&F_{(2,1)}=F_{(3,1)}=F_{(4,2)}=0\notag\\
&&F_{(4,1)},~F_{(5,1)},~F_{(5,2)}~ \mbox{are finite and at least one non-zero}.\notag\\
&&\label{class2_result}
\eea
As we mentioned earlier, the values of $F_i$'s at $X=0$ are not all independent. In fact, using eq. \eqref{F_relation} yields
\bea
F_{(4,1)}(X=0)&=&0,\notag\\
F_{(5,1)}(X=0)&=& - F_{(5,2)}(X=0)\neq 0.\label{class2_result2}
\eea
As a result, the only non-zero values among $F_i(X=0)$ are the ones coming from $G_5$.

In the derivation  above, we have only considered the limit to the flat metric through spherical perturbations, hence the above constitute a set of necessary conditions at this point. 
In the next section, we show that they are also sufficient condition for Class 2. Moreover, we discuss how Class 2 and Class 1 theories are related.

\subsection{Relation between Class 1 and 2}

Let us consider a Class 2 theory with Lagrangian $\L$. By the argument in the previous section, we know eqs. \eqref{class2_result} and \eqref{class2_result2} must hold.

Now let us define 
\beq
F_{(5,1)}(X)=c+\tilde F(X)
\eeq
where $c = F_{(5,1)}(X=0)$, thus $\tilde F(X=0)=0$. Using the definition of $F_{(5,1)}$ in terms of $G_5$, this gives
\beq\label{G_5_redef}
G_5=c \ln|X|+\int^XdX'\frac{\tilde F(X')}{X'}=c \ln|X|+\tilde G_5,
\eeq
where 
\beq
\tilde G_5\equiv \int^XdX'\frac{\tilde F(X')}{X'}
\eeq
satisfying $X\tilde G_{5X}=0$ at $X=0$.

By substituting $G_5$ from eq. \eqref{G_5_redef} in Lagrangian $\L$, the $\ln |X|$ contribution turns into a linear Gauss-Bonnet coupling term, and we get
\beq\label{Lagrangian_relation}
\L=\tilde \L-\frac{c}{4}\phi \mathcal{G}
\eeq
where $\tilde \L$ is a Lagrangian with $G_5$ in $\L$ is replaced by $\tilde G_5$\footnote{Note that Lagrangian $\L$ is linear in terms of $G_i$'s.}. Note that $\tilde \L$ satisfies eq. \eqref{class1_result}, thus it belongs to Class 1. Consequently, we have proven that the Lagrangian of any Class 2 theory is a linear Gauss-Bonnet coupling plus a Class 1 theory Lagrangian.

This result further proves that eqs. \eqref{class2_result} and \eqref{class2_result2} are {\it sufficient} conditions for Class 2 condition eq. \eqref{class2_condition} . In order to see this, consider the equation of motion derived from eq. \eqref{Lagrangian_relation} which reads
\beq
\E_\phi[\phi,\,g]=\tilde \E_\phi[\phi,\,g]-\frac{c}{4}\mathcal{G}.
\eeq
$\tilde \E_\phi$ is the equation of motion derived from $\tilde \L$ and satisfies $\tilde \E_\phi[\phi=0,\,g]=0$, as it belongs to Class 1. As a result,
\beq
\E_\phi[\phi=0,g]=-\frac{c}{4}\mathcal{G},
\eeq
which clearly satisfies eq. \eqref{class2_condition}. 

\section{Black hole solutions of Class 2}
So far, we have been working out the classification of SS Horndeski theories, and we have shown that Class 2 theories are closely related to the ones in Class 1. The scalar field for Class 2 theories {\it must} have a non-trivial configuration on curved spacetimes, and this means that their black hole solutions {\it must} have hair (This generalizes the logic of Ref.~\cite{Sotiriou:2013qea} to all theories in Class 2). We study the behaviour of the scalar hair in black hole spacetimes for Class 2 theories in this section.

It has been shown that for a canonical massless scalar field with linear Gauss-Bonnet coupling (corresponding to $\tilde \L= X$ in eq. \eqref{Lagrangian_relation}), the scalar hair of the black hole is secondary for stationary axi-symmetric \cite{Prabhu:2018aun} (static, spherically symmetric \cite{Benkel:2016rlz,Benkel:2016kcq}) black holes . In this section, we generalize this result to a wider class of theories with linear Gauss-Bonnet coupling. 

The first case that we consider is static, spherically symmetric spacetime and scalar field. The advantage in this case is that we do not put any restriction on the theory and asymptotics of the black hole. The proof applies to all Class 1 and Class 2 theories, {\em i.e.}~to all non-Lorentz breaking theories.

Then, we consider the case of a stationary, asymptotically flat black hole spacetime and stationary scalar field, without assuming extra spacetime symmetries. In this case, we assume that the dominant contribution to the kinetic term of the Lagrangian in the weak field limit is the canonical kinetic term. Essentially, this means that all corrections to the canonical kinetic term are {\it higher} order field contributions, that are relevant when the scalar field is strong. Let's place this condition in the context of a Class 2 theory with Lagrangian $\L$. As we have shown,  $\L$ can be expressed as  
\beq\label{GLSA_Lagrangian}
\L=\tilde \L +\alpha \phi \G 
\eeq
where $\tilde \L$ (with $\tilde G_i$ functions) is in Class 1 and $\alpha \neq 0$.\footnote{The arguments below apply to $\alpha = 0$ as well.} Hence, our condition regarding the weak field dominance of the canonical kinetic term implies that the dominant term to the Noether's current $\tilde J^a$ (of $\tilde \L$) at infinity is $-\nabla^a \phi$. Finally, we assume (for technical reasons of the proof) that the surface gravity of the Killing horizon is constant.

In both cases discussed above, and with the assumptions listed, we prove that the scalar charge is secondary.

\subsection{Spherical black holes}
Consider a static spherically symmetric black hole with the following metric
\beq\label{eq:coordinate}
g_{ab}dx^a dx^b=-h(r)dt^2 +\frac{dr^2}{f(r)}+r^2 d\Omega^2
\eeq
and a horizon at $r=r_H$ given by $f(r_H)=h(r_H)=0$. For a static spherically symmetric spacetime 
\beq
\G^a=(0,\G^r,0,0),\qquad \G^r=\frac{4\left(f-1\right)fh'}{r^2 h}.
\eeq
$J^a=\tilde J^a - \alpha \G^a$ satisfies the symmetries of the spacetime. Hence the only nontrivial component of $J^a$ in the  coordinates defined by eq. \eqref{eq:coordinate} is $J^r$. Explicitly solving $\nabla_a J^a=0$, we get
\beq
J^r= \frac{C}{r^2 }\sqrt{\frac{f}{h}}
\eeq
where $C$ is a constant. Thus, we have
\beq\label{eq:current_r_equation}
\frac{C}{r^2 }\sqrt{\frac{f}{h}}=\tilde J^r-\alpha\frac{4(f-1)fh'}{r^2 h}.
\eeq
The explicit form of $\tilde J^r$ is given by \cite{Sotiriou:2014pfa}
\bea
\tilde J^r&&=-f \phi' \tilde G_{2X}+f^2(\phi')^2\frac{rh'+4h}{2r h}\tilde G_{3X}\notag\\
&&+f\phi'\frac{2hf-2h+2rfh'}{r^2h}\tilde G_{4X}\notag\\
&&-2f^3 (\phi')^3\frac{h+rh'}{r^2h}\tilde G_{4XX}\notag\\
&&+f^2(\phi')^2h'\frac{1-3f}{2r^2 h}\tilde G_{5X}+f^4 (\phi')^4\frac{h'}{2r^2h}\tilde G_{5XX}.\label{Jr_explicit}
\eea
Assuming regularity of the scalar field at the horizon, {\em i.e.}~$\phi'(r_H)$ is finite, one can see  that $\tilde J^r(r_H)=0$ (in the next section, we will see the generalization of this result beyond spherical symmetry). As a result, evaluating eq.~\eqref{eq:current_r_equation} at $r=r_H$ fixes the constant $C$
\beq
C=4 \alpha \sqrt{f'(r_H)h'(r_H)}\, \mbox{sgn}(f'(r_H)).
\eeq

Substituting this value back in \eqref{eq:current_r_equation}, we get
\bea\label{eq:current_final}
\tilde J^r&=&\frac{4 \alpha}{r^2}\sqrt{\frac{f}{h}}\bigg[\sqrt{f'(r_H)h'(r_H)}\, \mbox{sgn}(f'(r_H))\notag\\
&&+\left(f-1\right)\sqrt{\frac{f}{h}}h'\bigg].
\eea
The important fact about this equation is that the right hand side depends only on the geometry and hence  $\tilde J^r$ is completely fixed by the spacetime.

The explicit form of $\tilde J^r$ in eq. \eqref{Jr_explicit} shows that it depends on the scalar field only through $\phi'$. In other words, combining eqs.~\eqref{eq:current_final} and \eqref{Jr_explicit} yields an algebraic equation for $\phi'$ in terms of geometrical quantities. 
This means that the scalar field configuration is completely fixed by the geometry. Hence,  the scalar hair is secondary and we have proven the desired result.
Note that in the above proof we did not use any restriction on the theory apart from not being Lorentz-violating ({\em i.e.}~not belonging in Class 3).

 In general, eq.~\eqref{eq:current_final} possesses multiple solutions for $\phi'$ even when $\alpha=0$. Treating the $\tilde G_i$'s as polynomial functions in $X$, we can re-arrange eq.~\eqref{eq:current_final} in the following format
\beq
\sum_{n=2}^{m} \alpha_nA_n(r)\phi'^n -f \phi' +\alpha A_0(r)=0,
\eeq
where $\alpha A_0(r)$ comes from the Gauss-Bonnet term [r.h.s of eq. \eqref{eq:current_final}], and the $\alpha_n A_n(r)$ terms originate from $\tilde J^r$, with $n$ controlled by the choice of the $\tilde G_i$.  We have assumed here that the canonical kinetic term is present and this gives the $f \phi'$ contribution. This polynomial equation can have multiple real roots. However, not all these roots will correspond to physically relevant solutions. 

To see this, first set $\alpha_n=0$, in which case one gets the unique (known) solution $\phi'=\alpha A_0(r)/f$. When $\alpha_n\neq 0$ this solution will receive corrections that vanish as $\alpha_n\to 0$ and the limit is smooth. More branches of solutions can also arise now, but these solutions are not expected to have a smooth limit as $\alpha_n\neq 0$ and hence they are not continuously connected with the only branch that exists for $\alpha_n=0$. Moreover, it is not clear whether any of these new solutions will have the correct asymptotic behaviour or whether they correspond to regular solutions of the whole theory (here only the scalar equation is considered). This behaviour persist for $\alpha= 0$, in which case $\alpha_n=0$ leads to the GR solution $\phi'=0$.

It is also worth pointing out that, for a spacetime with multiple horizons, $f'(r_H)$ changes sign on successive horizons. Hence, the value of $C$ calculated on different horizons (to ensure the regularity of $\phi$) cannot match. This means that the scalar field is singular on at least one of the horizons in such a spacetime. For example, spherically symmetric black holes with dS asymptotics are expected to be singular. This agrees with the results and conclusions obtained numerically in \cite{Brihaye:2017wln, Bakopoulos:2018nui}.

\subsection{Stationary black holes}
Now, let us present the proof for stationary black holes. The proof goes beyond spherical symmetry and uses only stationarity of the black hole spacetime. However, it is restricted to a subclass of theories in Class 1 and 2, as we have explained earlier. 

Consider a generic stationary asymptotically flat black hole spacetime with a Killing vector $\xi^a$ and Killing horizon $\H$. The equation of motion from Lagrangian \eqref{GLSA_Lagrangian} reads
\beq
\nabla_a \tilde J^a =\alpha \nabla_a \G^a 
\eeq
where $\tilde J^a$ is the Noether's current associated with $\tilde \L$ and $\nabla_a \G^a=\G$. Integrating above in a spacetime region bounded by the Killing horizon of the black hole ($\H$), infinity ($\infty$) and two (partial) Cauchy hypersurfaces ($C_1$ and $C_2$), we get
\beq
\int_\H n_a \tilde J^a +\int_\infty n_a \tilde J^a = \alpha \int_\H n_a \G^a +\alpha \int_\infty n_a \G^a,
\eeq
where $n_a$ is the normal to the boundary. Note that the integrals over $C_1$ and $C_2$ (by isometry) cancel each other.

On the Killing horizon $n_a=\xi_a$ and 
\beq
\xi_a \tilde J^a = 0,
\eeq 
provided that the scalar field is regular, static and $\H$ has  constant surface gravity \cite{Benkel:2018qmh}. Moreover,
\beq
\int_\infty n_a \G^a = 0
\eeq
for asymptotically flat spacetimes. 
Consequently,
\beq\label{scalar_charge}
\int_\infty n_a \tilde J^a = \alpha \int_\H n_a \G^a.
\eeq
The left hand side of the above gives the scalar charge of the black hole. In order to see this, consider ``$1/r$'' expansion of the scalar field near infinity,
\beq
\phi = \frac{C}{r}+\mathcal{O}(1/r^2).
\eeq
Substituting this expansion in the Noether's current (and imposing asymptotic flatness), we get
\beq\label{boundary_infinity}
\int_\infty n_a \tilde J^a = 4\pi C.
\eeq
Note that in above the only non-vanishing contribution is from $-\nabla^a \phi$ by the restriction we imposed on the theory that the canonical kinetic terms dominates in weak field. Substituting this back in eq. \eqref{scalar_charge} yeilds
\beq\label{scalar_charge_eq}
4\pi C = \alpha \int_\H n_a \G^a
\eeq
The right hand side of above is a purely geometrical quantity. As a result, the scalar charge is fixed by the geometry, {\em i.e.}~the scalar charge of the black hole is secondary. This generalizes the proof presented in Ref.~\cite{Prabhu:2018aun} beyond axisymmetry and to a wider class of theories within SS Horndeski; the charge of the scalar field is fixed by the properties of the horizon.

Eq. \eqref{scalar_charge_eq} holds for $\alpha =0$, corresponding to Class 1 theories too. In this case, we conclude that
\beq
C = 0.
\eeq 
In other words, the scalar charge of a hairy solution (if exists) of Class 1 theories vanishes, and the asymptotic fall off of the scalar field is faster than $1/r$.

We finish this section by the following observation. If the spacetime possesses multiple horizons, we expect eq. \eqref{scalar_charge_eq} to hold on each horizon. As a result (if $\alpha \neq 0$), 
\beq\label{horizon_matching}
\int_{\H_1} n_a \G^a=\int_{\H_2} n_a \G^a.
\eeq
This is a very restrictive condition on the spacetime geometry. In particular, the scalar field cannot remain regular if the above does not hold. As we have shown in the previous section, the above cannot hold for static spherically symmetric black holes. 
Eq. \eqref{horizon_matching} indicates that stationary black holes with multiple horizons (within theories we considered in this section) are also irregular at least on one of the horizons.

\section{Summary and Conclusion}
In this paper, we have presented a classification of shift-symmetric  Horndeski theories that could be useful in various applications. We argued that Horndeski theories can be split in three classes: (i) theories that admit all of the spacetimes of GR with a constant scalar configuration; (ii) theories do not belong to the previous class but that admit flat space with constant scalar; (iii) theories in which the scalar has to be nontrivial in flat space or do not admit flat space at all, and hence they are Lorentz-violating. We have identified the conditions on the $G_i$ function appearing in the action that correspond to each class of theories. We have also proven that the Lagranian of any theory in Class 2 is equal to the Lagrangian of some theory in Class 1 plus a term featuring a linear coupling between the scalar and the Gauss--Bonnet invariant. In particular, this means that any locally Lorentz invariant shift-symmetric Horndeski theory admits all GR solutions, provided it does not contain a linear Gauss-Bonnet coupling.

We have used our classifications to obtain some new results in  the context of no-hair theorems. In particular, we have shown that all theories in Class 2 will necessarily have hairy black hole solutions. We have further shown under fairly general conditions that the hair is secondary, {\em i.e.}~the scalar charge for these hairy black holes is fixed by the regularity of the horizon, and is determined by the horizon properties.

Our result underscore the  important role that a linear coupling between the  scalar field and Gauss-Bonnet term plays for black holes and complement the earlier results of Refs.~\cite{Sotiriou:2013qea,Sotiriou:2014pfa}. This is the unique interaction term that forces a Lorentz invariant theory within the shift-symmetric Horndeski class  to have hairy black hole solutions.

\begin{acknowledgments}
MS is supported by the Royal Commission for the Exhibition of 1851. TPS acknowledges partial support from the STFC Consolidated Grant No.~ST/P000703/1. We would also like to acknowledge network- ing support by the COST Action GWverse CA16104.

\end{acknowledgments}

\bibliography{Horndeski_classification}

\begin{thebibliography}{31}%
\makeatletter
\providecommand \@ifxundefined [1]{%
 \@ifx{#1\undefined}
}%
\providecommand \@ifnum [1]{%
 \ifnum #1\expandafter \@firstoftwo
 \else \expandafter \@secondoftwo
 \fi
}%
\providecommand \@ifx [1]{%
 \ifx #1\expandafter \@firstoftwo
 \else \expandafter \@secondoftwo
 \fi
}%
\providecommand \natexlab [1]{#1}%
\providecommand \enquote  [1]{``#1''}%
\providecommand \bibnamefont  [1]{#1}%
\providecommand \bibfnamefont [1]{#1}%
\providecommand \citenamefont [1]{#1}%
\providecommand \href@noop [0]{\@secondoftwo}%
\providecommand \href [0]{\begingroup \@sanitize@url \@href}%
\providecommand \@href[1]{\@@startlink{#1}\@@href}%
\providecommand \@@href[1]{\endgroup#1\@@endlink}%
\providecommand \@sanitize@url [0]{\catcode `\\12\catcode `\$12\catcode
  `\&12\catcode `\#12\catcode `\^12\catcode `\_12\catcode `\%12\relax}%
\providecommand \@@startlink[1]{}%
\providecommand \@@endlink[0]{}%
\providecommand \url  [0]{\begingroup\@sanitize@url \@url }%
\providecommand \@url [1]{\endgroup\@href {#1}{\urlprefix }}%
\providecommand \urlprefix  [0]{URL }%
\providecommand \Eprint [0]{\href }%
\providecommand \doibase [0]{http://dx.doi.org/}%
\providecommand \selectlanguage [0]{\@gobble}%
\providecommand \bibinfo  [0]{\@secondoftwo}%
\providecommand \bibfield  [0]{\@secondoftwo}%
\providecommand \translation [1]{[#1]}%
\providecommand \BibitemOpen [0]{}%
\providecommand \bibitemStop [0]{}%
\providecommand \bibitemNoStop [0]{.\EOS\space}%
\providecommand \EOS [0]{\spacefactor3000\relax}%
\providecommand \BibitemShut  [1]{\csname bibitem#1\endcsname}%
\let\auto@bib@innerbib\@empty
\bibitem [{\citenamefont {Lovelock}(1971)}]{Lovelock:1971yv}%
  \BibitemOpen
  \bibfield  {author} {\bibinfo {author} {\bibfnamefont {D.}~\bibnamefont
  {Lovelock}},\ }\href {\doibase 10.1063/1.1665613} {\bibfield  {journal}
  {\bibinfo  {journal} {J. Math. Phys.}\ }\textbf {\bibinfo {volume} {12}},\
  \bibinfo {pages} {498} (\bibinfo {year} {1971})}\BibitemShut {NoStop}%
\bibitem [{\citenamefont {Lovelock}(1972)}]{Lovelock:1972vz}%
  \BibitemOpen
  \bibfield  {author} {\bibinfo {author} {\bibfnamefont {D.}~\bibnamefont
  {Lovelock}},\ }\href {\doibase 10.1063/1.1666069} {\bibfield  {journal}
  {\bibinfo  {journal} {J. Math. Phys.}\ }\textbf {\bibinfo {volume} {13}},\
  \bibinfo {pages} {874} (\bibinfo {year} {1972})}\BibitemShut {NoStop}%
\bibitem [{\citenamefont {{Chase}}(1970)}]{1970CMaPh..19..276C}%
  \BibitemOpen
  \bibfield  {author} {\bibinfo {author} {\bibfnamefont {J.~E.}\ \bibnamefont
  {{Chase}}},\ }\href {\doibase 10.1007/BF01646635} {\bibfield  {journal}
  {\bibinfo  {journal} {Communications in Mathematical Physics}\ }\textbf
  {\bibinfo {volume} {19}},\ \bibinfo {pages} {276} (\bibinfo {year}
  {1970})}\BibitemShut {NoStop}%
\bibitem [{\citenamefont {Hawking}(1972)}]{Hawking:1972qk}%
  \BibitemOpen
  \bibfield  {author} {\bibinfo {author} {\bibfnamefont {S.~W.}\ \bibnamefont
  {Hawking}},\ }\href {\doibase 10.1007/BF01877518} {\bibfield  {journal}
  {\bibinfo  {journal} {Commun. Math. Phys.}\ }\textbf {\bibinfo {volume}
  {25}},\ \bibinfo {pages} {167} (\bibinfo {year} {1972})}\BibitemShut
  {NoStop}%
\bibitem [{\citenamefont {Bekenstein}(1995)}]{Bekenstein:1995un}%
  \BibitemOpen
  \bibfield  {author} {\bibinfo {author} {\bibfnamefont {J.~D.}\ \bibnamefont
  {Bekenstein}},\ }\href {\doibase 10.1103/PhysRevD.51.R6608} {\bibfield
  {journal} {\bibinfo  {journal} {Phys. Rev.}\ }\textbf {\bibinfo {volume}
  {D51}},\ \bibinfo {pages} {R6608} (\bibinfo {year} {1995})}\BibitemShut
  {NoStop}%
\bibitem [{\citenamefont {Sotiriou}\ and\ \citenamefont
  {Faraoni}(2012)}]{Sotiriou:2011dz}%
  \BibitemOpen
  \bibfield  {author} {\bibinfo {author} {\bibfnamefont {T.~P.}\ \bibnamefont
  {Sotiriou}}\ and\ \bibinfo {author} {\bibfnamefont {V.}~\bibnamefont
  {Faraoni}},\ }\href {\doibase 10.1103/PhysRevLett.108.081103} {\bibfield
  {journal} {\bibinfo  {journal} {Phys. Rev. Lett.}\ }\textbf {\bibinfo
  {volume} {108}},\ \bibinfo {pages} {081103} (\bibinfo {year} {2012})},\
  \Eprint {http://arxiv.org/abs/1109.6324} {arXiv:1109.6324 [gr-qc]}
  \BibitemShut {NoStop}%
\bibitem [{\citenamefont {Hui}\ and\ \citenamefont
  {Nicolis}(2013)}]{Hui:2012qt}%
  \BibitemOpen
  \bibfield  {author} {\bibinfo {author} {\bibfnamefont {L.}~\bibnamefont
  {Hui}}\ and\ \bibinfo {author} {\bibfnamefont {A.}~\bibnamefont {Nicolis}},\
  }\href {\doibase 10.1103/PhysRevLett.110.241104} {\bibfield  {journal}
  {\bibinfo  {journal} {Phys. Rev. Lett.}\ }\textbf {\bibinfo {volume} {110}},\
  \bibinfo {pages} {241104} (\bibinfo {year} {2013})},\ \Eprint
  {http://arxiv.org/abs/1202.1296} {arXiv:1202.1296 [hep-th]} \BibitemShut
  {NoStop}%
\bibitem [{\citenamefont {Sotiriou}\ and\ \citenamefont
  {Zhou}(2014{\natexlab{a}})}]{Sotiriou:2013qea}%
  \BibitemOpen
  \bibfield  {author} {\bibinfo {author} {\bibfnamefont {T.~P.}\ \bibnamefont
  {Sotiriou}}\ and\ \bibinfo {author} {\bibfnamefont {S.-Y.}\ \bibnamefont
  {Zhou}},\ }\href {\doibase 10.1103/PhysRevLett.112.251102} {\bibfield
  {journal} {\bibinfo  {journal} {Phys. Rev. Lett.}\ }\textbf {\bibinfo
  {volume} {112}},\ \bibinfo {pages} {251102} (\bibinfo {year}
  {2014}{\natexlab{a}})},\ \Eprint {http://arxiv.org/abs/1312.3622}
  {arXiv:1312.3622 [gr-qc]} \BibitemShut {NoStop}%
\bibitem [{\citenamefont {Sotiriou}(2015)}]{Sotiriou:2015pka}%
  \BibitemOpen
  \bibfield  {author} {\bibinfo {author} {\bibfnamefont {T.~P.}\ \bibnamefont
  {Sotiriou}},\ }\href {\doibase 10.1088/0264-9381/32/21/214002} {\bibfield
  {journal} {\bibinfo  {journal} {Class. Quant. Grav.}\ }\textbf {\bibinfo
  {volume} {32}},\ \bibinfo {pages} {214002} (\bibinfo {year} {2015})},\
  \Eprint {http://arxiv.org/abs/1505.00248} {arXiv:1505.00248 [gr-qc]}
  \BibitemShut {NoStop}%
\bibitem [{\citenamefont {Silva}\ \emph {et~al.}(2018)\citenamefont {Silva},
  \citenamefont {Sakstein}, \citenamefont {Gualtieri}, \citenamefont
  {Sotiriou},\ and\ \citenamefont {Berti}}]{Silva:2017uqg}%
  \BibitemOpen
  \bibfield  {author} {\bibinfo {author} {\bibfnamefont {H.~O.}\ \bibnamefont
  {Silva}}, \bibinfo {author} {\bibfnamefont {J.}~\bibnamefont {Sakstein}},
  \bibinfo {author} {\bibfnamefont {L.}~\bibnamefont {Gualtieri}}, \bibinfo
  {author} {\bibfnamefont {T.~P.}\ \bibnamefont {Sotiriou}}, \ and\ \bibinfo
  {author} {\bibfnamefont {E.}~\bibnamefont {Berti}},\ }\href {\doibase
  10.1103/PhysRevLett.120.131104} {\bibfield  {journal} {\bibinfo  {journal}
  {Phys. Rev. Lett.}\ }\textbf {\bibinfo {volume} {120}},\ \bibinfo {pages}
  {131104} (\bibinfo {year} {2018})},\ \Eprint
  {http://arxiv.org/abs/1711.02080} {arXiv:1711.02080 [gr-qc]} \BibitemShut
  {NoStop}%
\bibitem [{\citenamefont {Arvanitaki}\ and\ \citenamefont
  {Dubovsky}(2011)}]{Arvanitaki:2010sy}%
  \BibitemOpen
  \bibfield  {author} {\bibinfo {author} {\bibfnamefont {A.}~\bibnamefont
  {Arvanitaki}}\ and\ \bibinfo {author} {\bibfnamefont {S.}~\bibnamefont
  {Dubovsky}},\ }\href {\doibase 10.1103/PhysRevD.83.044026} {\bibfield
  {journal} {\bibinfo  {journal} {Phys. Rev.}\ }\textbf {\bibinfo {volume}
  {D83}},\ \bibinfo {pages} {044026} (\bibinfo {year} {2011})},\ \Eprint
  {http://arxiv.org/abs/1004.3558} {arXiv:1004.3558 [hep-th]} \BibitemShut
  {NoStop}%
\bibitem [{\citenamefont {Cardoso}\ \emph
  {et~al.}(2013{\natexlab{a}})\citenamefont {Cardoso}, \citenamefont {Carucci},
  \citenamefont {Pani},\ and\ \citenamefont {Sotiriou}}]{Cardoso:2013opa}%
  \BibitemOpen
  \bibfield  {author} {\bibinfo {author} {\bibfnamefont {V.}~\bibnamefont
  {Cardoso}}, \bibinfo {author} {\bibfnamefont {I.~P.}\ \bibnamefont
  {Carucci}}, \bibinfo {author} {\bibfnamefont {P.}~\bibnamefont {Pani}}, \
  and\ \bibinfo {author} {\bibfnamefont {T.~P.}\ \bibnamefont {Sotiriou}},\
  }\href {\doibase 10.1103/PhysRevD.88.044056} {\bibfield  {journal} {\bibinfo
  {journal} {Phys. Rev.}\ }\textbf {\bibinfo {volume} {D88}},\ \bibinfo {pages}
  {044056} (\bibinfo {year} {2013}{\natexlab{a}})},\ \Eprint
  {http://arxiv.org/abs/1305.6936} {arXiv:1305.6936 [gr-qc]} \BibitemShut
  {NoStop}%
\bibitem [{\citenamefont {Cardoso}\ \emph
  {et~al.}(2013{\natexlab{b}})\citenamefont {Cardoso}, \citenamefont {Carucci},
  \citenamefont {Pani},\ and\ \citenamefont {Sotiriou}}]{Cardoso:2013fwa}%
  \BibitemOpen
  \bibfield  {author} {\bibinfo {author} {\bibfnamefont {V.}~\bibnamefont
  {Cardoso}}, \bibinfo {author} {\bibfnamefont {I.~P.}\ \bibnamefont
  {Carucci}}, \bibinfo {author} {\bibfnamefont {P.}~\bibnamefont {Pani}}, \
  and\ \bibinfo {author} {\bibfnamefont {T.~P.}\ \bibnamefont {Sotiriou}},\
  }\href {\doibase 10.1103/PhysRevLett.111.111101} {\bibfield  {journal}
  {\bibinfo  {journal} {Phys. Rev. Lett.}\ }\textbf {\bibinfo {volume} {111}},\
  \bibinfo {pages} {111101} (\bibinfo {year} {2013}{\natexlab{b}})},\ \Eprint
  {http://arxiv.org/abs/1308.6587} {arXiv:1308.6587 [gr-qc]} \BibitemShut
  {NoStop}%
\bibitem [{\citenamefont {Babichev}\ and\ \citenamefont
  {Charmousis}(2014)}]{Babichev:2013cya}%
  \BibitemOpen
  \bibfield  {author} {\bibinfo {author} {\bibfnamefont {E.}~\bibnamefont
  {Babichev}}\ and\ \bibinfo {author} {\bibfnamefont {C.}~\bibnamefont
  {Charmousis}},\ }\href {\doibase 10.1007/JHEP08(2014)106} {\bibfield
  {journal} {\bibinfo  {journal} {JHEP}\ }\textbf {\bibinfo {volume} {08}},\
  \bibinfo {pages} {106} (\bibinfo {year} {2014})},\ \Eprint
  {http://arxiv.org/abs/1312.3204} {arXiv:1312.3204 [gr-qc]} \BibitemShut
  {NoStop}%
\bibitem [{\citenamefont {Herdeiro}\ and\ \citenamefont
  {Radu}(2014)}]{Herdeiro:2014goa}%
  \BibitemOpen
  \bibfield  {author} {\bibinfo {author} {\bibfnamefont {C.~A.~R.}\
  \bibnamefont {Herdeiro}}\ and\ \bibinfo {author} {\bibfnamefont
  {E.}~\bibnamefont {Radu}},\ }\href {\doibase 10.1103/PhysRevLett.112.221101}
  {\bibfield  {journal} {\bibinfo  {journal} {Phys. Rev. Lett.}\ }\textbf
  {\bibinfo {volume} {112}},\ \bibinfo {pages} {221101} (\bibinfo {year}
  {2014})},\ \Eprint {http://arxiv.org/abs/1403.2757} {arXiv:1403.2757 [gr-qc]}
  \BibitemShut {NoStop}%
\bibitem [{\citenamefont {Babichev}\ \emph {et~al.}(2016)\citenamefont
  {Babichev}, \citenamefont {Charmousis},\ and\ \citenamefont
  {Lehébel}}]{Babichev:2016rlq}%
  \BibitemOpen
  \bibfield  {author} {\bibinfo {author} {\bibfnamefont {E.}~\bibnamefont
  {Babichev}}, \bibinfo {author} {\bibfnamefont {C.}~\bibnamefont
  {Charmousis}}, \ and\ \bibinfo {author} {\bibfnamefont {A.}~\bibnamefont
  {Lehébel}},\ }\href {\doibase 10.1088/0264-9381/33/15/154002} {\bibfield
  {journal} {\bibinfo  {journal} {Class. Quant. Grav.}\ }\textbf {\bibinfo
  {volume} {33}},\ \bibinfo {pages} {154002} (\bibinfo {year} {2016})},\
  \Eprint {http://arxiv.org/abs/1604.06402} {arXiv:1604.06402 [gr-qc]}
  \BibitemShut {NoStop}%
\bibitem [{\citenamefont {Doneva}\ and\ \citenamefont
  {Yazadjiev}(2018)}]{Doneva:2017bvd}%
  \BibitemOpen
  \bibfield  {author} {\bibinfo {author} {\bibfnamefont {D.~D.}\ \bibnamefont
  {Doneva}}\ and\ \bibinfo {author} {\bibfnamefont {S.~S.}\ \bibnamefont
  {Yazadjiev}},\ }\href {\doibase 10.1103/PhysRevLett.120.131103} {\bibfield
  {journal} {\bibinfo  {journal} {Phys. Rev. Lett.}\ }\textbf {\bibinfo
  {volume} {120}},\ \bibinfo {pages} {131103} (\bibinfo {year} {2018})},\
  \Eprint {http://arxiv.org/abs/1711.01187} {arXiv:1711.01187 [gr-qc]}
  \BibitemShut {NoStop}%
\bibitem [{\citenamefont {Antoniou}\ \emph {et~al.}(2018)\citenamefont
  {Antoniou}, \citenamefont {Bakopoulos},\ and\ \citenamefont
  {Kanti}}]{Antoniou:2017acq}%
  \BibitemOpen
  \bibfield  {author} {\bibinfo {author} {\bibfnamefont {G.}~\bibnamefont
  {Antoniou}}, \bibinfo {author} {\bibfnamefont {A.}~\bibnamefont
  {Bakopoulos}}, \ and\ \bibinfo {author} {\bibfnamefont {P.}~\bibnamefont
  {Kanti}},\ }\href {\doibase 10.1103/PhysRevLett.120.131102} {\bibfield
  {journal} {\bibinfo  {journal} {Phys. Rev. Lett.}\ }\textbf {\bibinfo
  {volume} {120}},\ \bibinfo {pages} {131102} (\bibinfo {year} {2018})},\
  \Eprint {http://arxiv.org/abs/1711.03390} {arXiv:1711.03390 [hep-th]}
  \BibitemShut {NoStop}%
\bibitem [{\citenamefont {Kanti}\ \emph {et~al.}(1996)\citenamefont {Kanti},
  \citenamefont {Mavromatos}, \citenamefont {Rizos}, \citenamefont {Tamvakis},\
  and\ \citenamefont {Winstanley}}]{Kanti:1995vq}%
  \BibitemOpen
  \bibfield  {author} {\bibinfo {author} {\bibfnamefont {P.}~\bibnamefont
  {Kanti}}, \bibinfo {author} {\bibfnamefont {N.~E.}\ \bibnamefont
  {Mavromatos}}, \bibinfo {author} {\bibfnamefont {J.}~\bibnamefont {Rizos}},
  \bibinfo {author} {\bibfnamefont {K.}~\bibnamefont {Tamvakis}}, \ and\
  \bibinfo {author} {\bibfnamefont {E.}~\bibnamefont {Winstanley}},\ }\href
  {\doibase 10.1103/PhysRevD.54.5049} {\bibfield  {journal} {\bibinfo
  {journal} {Phys. Rev.}\ }\textbf {\bibinfo {volume} {D54}},\ \bibinfo {pages}
  {5049} (\bibinfo {year} {1996})},\ \Eprint
  {http://arxiv.org/abs/hep-th/9511071} {arXiv:hep-th/9511071 [hep-th]}
  \BibitemShut {NoStop}%
\bibitem [{\citenamefont {Yunes}\ and\ \citenamefont
  {Stein}(2011)}]{Yunes:2011we}%
  \BibitemOpen
  \bibfield  {author} {\bibinfo {author} {\bibfnamefont {N.}~\bibnamefont
  {Yunes}}\ and\ \bibinfo {author} {\bibfnamefont {L.~C.}\ \bibnamefont
  {Stein}},\ }\href {\doibase 10.1103/PhysRevD.83.104002} {\bibfield  {journal}
  {\bibinfo  {journal} {Phys. Rev.}\ }\textbf {\bibinfo {volume} {D83}},\
  \bibinfo {pages} {104002} (\bibinfo {year} {2011})},\ \Eprint
  {http://arxiv.org/abs/1101.2921} {arXiv:1101.2921 [gr-qc]} \BibitemShut
  {NoStop}%
\bibitem [{\citenamefont {{Horndeski}}(1974)}]{1974IJTP...10..363H}%
  \BibitemOpen
  \bibfield  {author} {\bibinfo {author} {\bibfnamefont {G.~W.}\ \bibnamefont
  {{Horndeski}}},\ }\href {\doibase 10.1007/BF01807638} {\bibfield  {journal}
  {\bibinfo  {journal} {International Journal of Theoretical Physics}\ }\textbf
  {\bibinfo {volume} {10}},\ \bibinfo {pages} {363} (\bibinfo {year}
  {1974})}\BibitemShut {NoStop}%
\bibitem [{\citenamefont {Deffayet}\ \emph {et~al.}(2011)\citenamefont
  {Deffayet}, \citenamefont {Gao}, \citenamefont {Steer},\ and\ \citenamefont
  {Zahariade}}]{Deffayet:2011gz}%
  \BibitemOpen
  \bibfield  {author} {\bibinfo {author} {\bibfnamefont {C.}~\bibnamefont
  {Deffayet}}, \bibinfo {author} {\bibfnamefont {X.}~\bibnamefont {Gao}},
  \bibinfo {author} {\bibfnamefont {D.~A.}\ \bibnamefont {Steer}}, \ and\
  \bibinfo {author} {\bibfnamefont {G.}~\bibnamefont {Zahariade}},\ }\href
  {\doibase 10.1103/PhysRevD.84.064039} {\bibfield  {journal} {\bibinfo
  {journal} {Phys. Rev.}\ }\textbf {\bibinfo {volume} {D84}},\ \bibinfo {pages}
  {064039} (\bibinfo {year} {2011})},\ \Eprint {http://arxiv.org/abs/1103.3260}
  {arXiv:1103.3260 [hep-th]} \BibitemShut {NoStop}%
\bibitem [{\citenamefont {Kobayashi}\ \emph {et~al.}(2011)\citenamefont
  {Kobayashi}, \citenamefont {Yamaguchi},\ and\ \citenamefont
  {Yokoyama}}]{Kobayashi:2011nu}%
  \BibitemOpen
  \bibfield  {author} {\bibinfo {author} {\bibfnamefont {T.}~\bibnamefont
  {Kobayashi}}, \bibinfo {author} {\bibfnamefont {M.}~\bibnamefont
  {Yamaguchi}}, \ and\ \bibinfo {author} {\bibfnamefont {J.}~\bibnamefont
  {Yokoyama}},\ }\href {\doibase 10.1143/PTP.126.511} {\bibfield  {journal}
  {\bibinfo  {journal} {Prog. Theor. Phys.}\ }\textbf {\bibinfo {volume}
  {126}},\ \bibinfo {pages} {511} (\bibinfo {year} {2011})},\ \Eprint
  {http://arxiv.org/abs/1105.5723} {arXiv:1105.5723 [hep-th]} \BibitemShut
  {NoStop}%
\bibitem [{\citenamefont {Sotiriou}\ and\ \citenamefont
  {Zhou}(2014{\natexlab{b}})}]{Sotiriou:2014pfa}%
  \BibitemOpen
  \bibfield  {author} {\bibinfo {author} {\bibfnamefont {T.~P.}\ \bibnamefont
  {Sotiriou}}\ and\ \bibinfo {author} {\bibfnamefont {S.-Y.}\ \bibnamefont
  {Zhou}},\ }\href {\doibase 10.1103/PhysRevD.90.124063} {\bibfield  {journal}
  {\bibinfo  {journal} {Phys. Rev.}\ }\textbf {\bibinfo {volume} {D90}},\
  \bibinfo {pages} {124063} (\bibinfo {year} {2014}{\natexlab{b}})},\ \Eprint
  {http://arxiv.org/abs/1408.1698} {arXiv:1408.1698 [gr-qc]} \BibitemShut
  {NoStop}%
\bibitem [{\citenamefont {Afshordi}\ \emph {et~al.}(2007)\citenamefont
  {Afshordi}, \citenamefont {Chung},\ and\ \citenamefont
  {Geshnizjani}}]{Afshordi:2006ad}%
  \BibitemOpen
  \bibfield  {author} {\bibinfo {author} {\bibfnamefont {N.}~\bibnamefont
  {Afshordi}}, \bibinfo {author} {\bibfnamefont {D.~J.~H.}\ \bibnamefont
  {Chung}}, \ and\ \bibinfo {author} {\bibfnamefont {G.}~\bibnamefont
  {Geshnizjani}},\ }\href {\doibase 10.1103/PhysRevD.75.083513} {\bibfield
  {journal} {\bibinfo  {journal} {Phys. Rev.}\ }\textbf {\bibinfo {volume}
  {D75}},\ \bibinfo {pages} {083513} (\bibinfo {year} {2007})},\ \Eprint
  {http://arxiv.org/abs/hep-th/0609150} {arXiv:hep-th/0609150 [hep-th]}
  \BibitemShut {NoStop}%
\bibitem [{\citenamefont {Prabhu}\ and\ \citenamefont
  {Stein}(2018)}]{Prabhu:2018aun}%
  \BibitemOpen
  \bibfield  {author} {\bibinfo {author} {\bibfnamefont {K.}~\bibnamefont
  {Prabhu}}\ and\ \bibinfo {author} {\bibfnamefont {L.~C.}\ \bibnamefont
  {Stein}},\ }\href {\doibase 10.1103/PhysRevD.98.021503} {\bibfield  {journal}
  {\bibinfo  {journal} {Phys. Rev.}\ }\textbf {\bibinfo {volume} {D98}},\
  \bibinfo {pages} {021503} (\bibinfo {year} {2018})},\ \Eprint
  {http://arxiv.org/abs/1805.02668} {arXiv:1805.02668 [gr-qc]} \BibitemShut
  {NoStop}%
\bibitem [{\citenamefont {Benkel}\ \emph {et~al.}(2017)\citenamefont {Benkel},
  \citenamefont {Sotiriou},\ and\ \citenamefont {Witek}}]{Benkel:2016rlz}%
  \BibitemOpen
  \bibfield  {author} {\bibinfo {author} {\bibfnamefont {R.}~\bibnamefont
  {Benkel}}, \bibinfo {author} {\bibfnamefont {T.~P.}\ \bibnamefont
  {Sotiriou}}, \ and\ \bibinfo {author} {\bibfnamefont {H.}~\bibnamefont
  {Witek}},\ }\href {\doibase 10.1088/1361-6382/aa5ce7} {\bibfield  {journal}
  {\bibinfo  {journal} {Class. Quant. Grav.}\ }\textbf {\bibinfo {volume}
  {34}},\ \bibinfo {pages} {064001} (\bibinfo {year} {2017})},\ \Eprint
  {http://arxiv.org/abs/1610.09168} {arXiv:1610.09168 [gr-qc]} \BibitemShut
  {NoStop}%
\bibitem [{\citenamefont {Benkel}\ \emph {et~al.}(2016)\citenamefont {Benkel},
  \citenamefont {Sotiriou},\ and\ \citenamefont {Witek}}]{Benkel:2016kcq}%
  \BibitemOpen
  \bibfield  {author} {\bibinfo {author} {\bibfnamefont {R.}~\bibnamefont
  {Benkel}}, \bibinfo {author} {\bibfnamefont {T.~P.}\ \bibnamefont
  {Sotiriou}}, \ and\ \bibinfo {author} {\bibfnamefont {H.}~\bibnamefont
  {Witek}},\ }\href {\doibase 10.1103/PhysRevD.94.121503} {\bibfield  {journal}
  {\bibinfo  {journal} {Phys. Rev.}\ }\textbf {\bibinfo {volume} {D94}},\
  \bibinfo {pages} {121503} (\bibinfo {year} {2016})},\ \Eprint
  {http://arxiv.org/abs/1612.08184} {arXiv:1612.08184 [gr-qc]} \BibitemShut
  {NoStop}%
\bibitem [{\citenamefont {Brihaye}\ \emph {et~al.}(2018)\citenamefont
  {Brihaye}, \citenamefont {Hartmann},\ and\ \citenamefont
  {Urrestilla}}]{Brihaye:2017wln}%
  \BibitemOpen
  \bibfield  {author} {\bibinfo {author} {\bibfnamefont {Y.}~\bibnamefont
  {Brihaye}}, \bibinfo {author} {\bibfnamefont {B.}~\bibnamefont {Hartmann}}, \
  and\ \bibinfo {author} {\bibfnamefont {J.}~\bibnamefont {Urrestilla}},\
  }\href {\doibase 10.1007/JHEP06(2018)074} {\bibfield  {journal} {\bibinfo
  {journal} {JHEP}\ }\textbf {\bibinfo {volume} {06}},\ \bibinfo {pages} {074}
  (\bibinfo {year} {2018})},\ \Eprint {http://arxiv.org/abs/1712.02458}
  {arXiv:1712.02458 [gr-qc]} \BibitemShut {NoStop}%
\bibitem [{\citenamefont {Bakopoulos}\ \emph {et~al.}(2018)\citenamefont
  {Bakopoulos}, \citenamefont {Antoniou},\ and\ \citenamefont
  {Kanti}}]{Bakopoulos:2018nui}%
  \BibitemOpen
  \bibfield  {author} {\bibinfo {author} {\bibfnamefont {A.}~\bibnamefont
  {Bakopoulos}}, \bibinfo {author} {\bibfnamefont {G.}~\bibnamefont
  {Antoniou}}, \ and\ \bibinfo {author} {\bibfnamefont {P.}~\bibnamefont
  {Kanti}},\ }\href@noop {} {\  (\bibinfo {year} {2018})},\ \Eprint
  {http://arxiv.org/abs/1812.06941} {arXiv:1812.06941 [hep-th]} \BibitemShut
  {NoStop}%
\bibitem [{\citenamefont {Benkel}\ \emph {et~al.}(2018)\citenamefont {Benkel},
  \citenamefont {Franchini}, \citenamefont {Saravani},\ and\ \citenamefont
  {Sotiriou}}]{Benkel:2018qmh}%
  \BibitemOpen
  \bibfield  {author} {\bibinfo {author} {\bibfnamefont {R.}~\bibnamefont
  {Benkel}}, \bibinfo {author} {\bibfnamefont {N.}~\bibnamefont {Franchini}},
  \bibinfo {author} {\bibfnamefont {M.}~\bibnamefont {Saravani}}, \ and\
  \bibinfo {author} {\bibfnamefont {T.~P.}\ \bibnamefont {Sotiriou}},\
  }\href@noop {} {\  (\bibinfo {year} {2018})},\ \Eprint
  {http://arxiv.org/abs/1806.08214} {arXiv:1806.08214 [gr-qc]} \BibitemShut
  {NoStop}%
\end{thebibliography}%
\end{document}